\documentclass[aps,pre,amsmath,amssymb,reprint,superscriptaddress]{revtex4-1}
\usepackage{graphicx}
\usepackage{dcolumn}
\usepackage{bm}

\usepackage[utf8]{inputenc}
\usepackage[T1]{fontenc}
\usepackage{mathptmx}
\usepackage{textcomp}
\usepackage{mathtools}

\newcommand{\IFUFG}{Instituto de F{\'i}sica, Universidade Federal de 
Goi{\'a}s, Av. Esperan\c{c}a s/n, 74.690-900, Goi{\^a}nia, GO, Brazil}
\newcommand{\IFMT}{Instituto Federal do Mato Grosso - Campus C{\'a}ceres,
Av. dos Ramires s/n, 78200-000, C{\'a}ceres, MT, Brazil}
\newcommand{\IFG}{Instituto Federal de Goi{\'a}s, Rua 76, Centro, Goi{\^a}nia - GO, Brazil}

\begin{document}


\title{Thoroughly analysis of the phase diagram for the Bell-Lavis model: \\An entropic simulational study} 


\author{L. S. Ferreira}
\email{ferreira_s_lucas@ufg.br}
\affiliation{\IFUFG}
\author{L. N. Jorge}
\affiliation{\IFMT}
\author{Cl\'{a}udio J. DaSilva}
\affiliation{\IFG}
\author{A. A. Caparica}
\affiliation{\IFUFG}
\email{caparica@ufg.br}


\date{\today}

\begin{abstract}
In this work, we investigate the Bell-Lavis model using entropic simulations for several values of the energy parameters. The $T\times\mu$ phase diagram and the ground state configurations are analyzed thoroughly. Besides, we examine the particle density and specific heat behavior for different values of the chemical potential $\mu$ as functions of temperature. We also obtain configurations that maximize the canonical probability for several values of chemical potential and temperature, enabling the identification of the low density ($LDL$) and high-density liquid ($HDL$) phases, among others, in the critical regions. We found a second-order phase transition from the $LDL-HDL_0$ to $LDL-HDL$ coexistence in the range of $0<\mu<1.05503$. In the $1.05503<\mu<1.48024$ range, the transition between the $LDL-HDL_0$ and $LDL-HDL_0-empty$ coexistence presents discontinuous and continuous transitions characteristics. Finally, for $1.48024<\mu<1.5$, the phase transition between $LDL$ and $empty$ phases is of first-order.
\end{abstract}


\maketitle 

\section{Introduction}


A phase transition can be considered a structural change in the macroscopic state of a thermodynamic system through changes in the external parameters \cite{Stanley1971, Landau1980, Goldenfeld1992}. Theoretically, they can be identified as abrupt variations in thermodynamic quantities of the system, such as energy, specific heat, order parameter, and susceptibility of the order parameter. However, looking only at these quantities, it is not possible to identify which configurations are present in the phase transition region.

A classical example of such a situation emerges from the analysis of a magnetic system described by the two-dimensional (2D) Ising model \cite{Ising1925} in the absence of an external magnetic field. It is well-known that for low temperatures, this model is in an ordered state, with all spins up or down, and as the temperature increases, the system undergoes a phase transition to the disordered phase evidenced by the value of the magnetization per site going from one to zero. Commonly, the ordered-phase is well characterized, while the disordered one has a high degree of degeneration, which includes configurations where the spins assemble magnetic domains and paramagnetic structures. Hence, it is clear that the magnetization itself does not allow ones to distinguish these structures, having the same value in both cases. Failure to know to which state the system undergoes a transition may lead to disagreement between experiment and theoretical approaches. For the 2D Ising model case, the system passes through configurations with magnetic domains when $T \sim T_c$ and to paramagnetic configurations for $T>T_c$ \cite{Christensen2005, Landau2014}, where $T_c$ is the critical temperature.

One way to circumvent this type of situation is to observe the most probable configurations for a given temperature. If the density of states of the system is known, we can calculate the partition function and the canonical probability given by
\begin{equation}\label{probability}
 P(E,T)=\frac{g(E)e^{-\beta E}}{Z},
\end{equation}
where $Z$ is the partition function given by 
\begin{equation}
 Z=\sum_{E'} g(E')e^{-\beta E'},
\end{equation}
where $\beta=1/k_BT$, $k_B$ is the Boltzmann's constant, $T$ is the absolute temperature, and $E'$ runs over all possible energies of the system. 

One way to obtain the density of states is to carry out an entropic Monte Carlo simulation \cite{Wang2001, Wang2001a, Caparica2012, Caparica2014, Jorge2016, Ferreira2019}, that have proven to be a powerful tool to the study of phase transitions. With the canonical probability, one can obtain the energy that maximizes it at a given temperature and thereby filter the configurations that satisfy the maximization condition, having an overview of the configurations present at each temperature and especially at the critical temperature \cite{Jorge2019, Nomura2017}.

In its turn, the comprehension of the thermodynamic properties of water is a great challenge addressed both experimentally and theoretically. It exhibits several different behaviors from other substances, one of which is the maximum in the particle density around 4 $^\circ$C \cite{Mishima1998, Limmer2013, Perakis2017}. Theoretically, several models were proposed, which include continuous \cite{Jagla1998, Sadr-Lahijany1999, Scala2001, Franzese2001, Gibson2006} and lattice models
\cite{Henriques2005, Truskett2002, Somer1995, Silverstein1999}, which allow accessing regions of the configurational space unattainable in experiments. One of the simplest models to represent water was proposed by Bell and Lavis 
\cite{Lavis1973}, which takes into account the orientational characteristic of hydrogen bonds and permits observing the maximum particle density. 

The Bell-Lavis (BL) model was already studied using mean-field approximations 
\cite {Bell1970, Lavis1973}, renormalization group analysis 
\cite{Young1979, Southern1980}, cluster variation method 
\cite{Buzano2008} and Bethe calculations for Husimi cactus 
\cite{Barbosa2008}. Also, several authors used Monte Carlo simulations in this context \cite{Fiore2009, Simenas2014, Simenas2015}. These works show that in the region of the maximum particle density, the system is in a transition between two phases known as high-density liquid ($HDL$) and low-density liquid ($LDL$). Nonetheless, the proper order parameter required to characterize these two phases is still missing, which makes impossible the full comprehension of this transition.

In view of the challenge of understanding the transition between the $LDL$ and $HDL$ phases in the Bell-Lavis model, we use
entropic simulations to obtain the thermodynamic properties of this system, as well as the configurations that maximize the canonical probability at the critical temperature, in order to characterize the phases through which the system visits during the phase transition. By means of this study we intend to obtain the necessary information to propose an order parameter for the transition between the phases $LDL$ and $HDL$.

\section{Bell-Lavis Model}
 
\begin{figure}[t]
\centering
\includegraphics[scale=1]{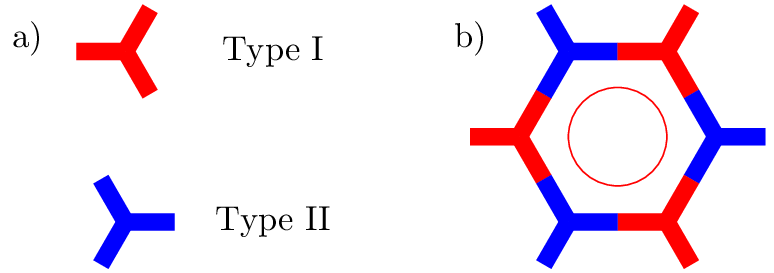}
\caption{a) The two types of molecules that can be in the lattice. b) Example of a $LDL$ configuration. \label{fig1}}
\end{figure}

In 1970, Bell and Lavis \cite{Bell1970} proposed a two-dimensional model for liquid systems that takes into account the orientational characteristic of hydrogen bonds in a triangular lattice. The lattice components are trimers whose interaction arms are 
120\textdegree \hspace{0.05cm} from each other, so that there are only two types of molecules, as can be seen in figure \ref{fig1} a). Interactions between first neighbors can occur in two ways: van der Waals potential or hydrogen bonding, the latter type only occurring if the arms of the neighboring trimers point towards each other, as shown in figure \ref{fig1} b). Each site occupied by a molecule has only three possibilities of performing hydrogen bonding interactions and directions, depending on its type. The open circle represents the absence of a particle at the site. Such representation is rather pictorial when compared to the real situation. Nevertheless, the physical characteristics to be considered here are easily attained using this model.

The Hamiltonian of the system can be written as
\begin{equation}\label{eq:bl}
 \mathcal{H}=-\sum_{\langle i,j\rangle} \eta_{i}\eta_{j}\left(\epsilon_{vdW}+\epsilon_{H}\tau_i^{ij}\tau_j^{ji}\right)+\mu \sum_i \eta_i,
\end{equation}
where $\eta_i$ is a site occupation variable with $1(0)$ being occupied (empty). $\epsilon_{vdW}$ and $\epsilon_{H}$ are the energetic parameters of the two types of interaction. The
$\tau_i^{ij}$ and $\tau_j^{ji}$ terms are responsible for the interaction directions of trimers, with $\tau_i^{ij}=1(0)$ in case the molecule at the $i$ site has an arm that points to 
$j$(otherwise), similarly to $\tau_j^{ji}$. The last term represents the interaction due to the chemical potential $\mu$. The first sum runs over the first neighbors and the last over all sites in the lattice.
\begin{figure}[t]
\includegraphics[scale=1.5]{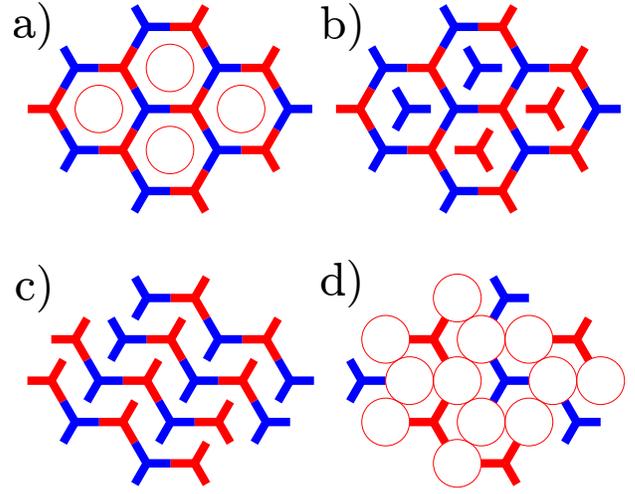}
\caption{Some phases presented by the Bell-Lavis model. In a) $LDL$ configuration, b) $HDL_0$ configuration, c) $HDL$ configuration and d) a possible interleaved phase.\label{fig2}}
\end{figure}

Fig. \ref{fig2} shows some phases of the system (ordered configurations). The phase represented in Fig. \ref{fig2} a) is known as a low density liquid ($LDL$). The phases \ref{fig2} b) and \ref{fig2} c) are called high density liquid ($HDL$), and we will name the phase in Fig. \ref{fig2} b) $HDL_0$. We have shown only two of these configurations, but there are several others, for example, in Fig. \ref{fig2} b) the centers of the hexagons possess molecules of type I and II, but they can be type I or II alone. Fig. \ref{fig2} d) shows a possible phase without interaction. An illustration of how these phases are arranged, according to energy parameters, was obtained by Barbosa and Henriques \cite{Barbosa2008}.

To characterize the different phases, we will use the particle density given by
\begin{equation}
 \rho=\frac{1}{L^2}\sum_{i,j} (\delta_{1,m_{i,j}}+\delta_{2,m_{i,j}}),
\end{equation}
where $m_{i,j}$ is the type of molecule that occupies the $i,j$ site and the first index refers to the type of molecule.
In particular, we can define the type I and II particle densities, respectively, as
\begin{align}
 n_1=&\frac{1}{L^2}\sum_{i,j} \delta_{1,m_{i,j}}\\
 n_2=&\frac{1}{L^2}\sum_{i,j} \delta_{2,m_{i,j}}.
\end{align}
The density of hydrogen bonds is obtained by
\begin{equation}
 e_{H}=\frac{1}{L^2}\sum_{\langle i,j\rangle} \eta_{i}\eta_{j}\tau_i^{ij}\tau_j^{ji}.
\end{equation}

In this model, we have three energy parameters $\epsilon_{vdW}$ - van der Waals interaction energy, $\epsilon_{H}$ - hydrogen bonding energy, and $\mu$ - chemical potential. In this work, we are interested in the transition between the $LDL$ and $HDL$ phases that occur for various values of the energy parameters as represented by the diagram of the reference \cite{Barbosa2008}, in particular $\epsilon_{vdW}=0$. In this situation, with the variation of the chemical potential, we can obtain the following ordered configurations: empty lattice, $LDL$, and $HDL$. We will restrict our study to the case $\epsilon_{vdW}=0$ so that we can rewrite the Hamiltonian as follows
\begin{equation}
\mathcal{H}=-\epsilon_{H}E_H + \mu N,
\end{equation}
where
\begin{equation}
 E_H=\sum_{i,j} \eta_{i}\eta_{j}\tau_i^{ij}\tau_j^{ji}
\end{equation}
and
\begin{equation}
 N=\sum_{i,j} (\delta_{1,n_{i,j}}+\delta_{2,n_{i,j}})
\end{equation}
are the number of hydrogen bonds and the number of particles, respectively.

\section{Computational Details}

To obtain the mean values of the quantities that characterize the system, we carry out entropic simulations based on the Wang-Landau method\cite{Wang2001}, which allows the estimation of the joint density of states, making it possible to obtain canonical averages as functions of temperature and energetic variables
\cite{Ferreira2019}. This methodology has not yet been used to investigate the Bell-Lavis model.

To perform the simulations, we use an isomorphic transformation from a triangular lattice into a square one of $L\times L$ sites \cite{Jorge2019}. The lattice sites can be in three possible states: 1 - type I molecule, 2 - type II molecule, or 3 - empty. We only count the hydrogen bond when two near-neighboring sites are in different states 1 and 2. And the number of particles is the sum of the number of sites in states 1 and 2.

We perform entropic simulations using the joint density of states with two parameters $g(E_{H}, N)$ related to the interaction energies of the system. The simulation procedure is as follows. An attempt to change the configuration of the system is made by changing the state of one site, chosen randomly, to any other state, being accepted with a probability equal to
\begin{equation}
p(E_H, N \rightarrow E_H', N')=\text{min}\left\{\frac{g(E_H, N)}{g(E_H', N')},1\right\}
\end{equation}  
where $E_H, N$, and $E_H', N'$ are the hydrogen bond and particle numbers of the current configuration and the tentative configuration, respectively. The density of states and the energy histogram are updated after a Monte Carlo step, that is, after $L^2$ attempted changes \cite{Caparica2012}. The histogram is flat only if $H(E_H, N)>0.8 \langle H \rangle$ for all hydrogen bond numbers and particle numbers, where $\langle H \rangle $ is the average over all hydrogen bonds and particles. The microcanonical averages of the particle densities of each molecule type are stored only after the seventh WL level\cite{Caparica2012}. The simulation ends when it reaches the sixteenth WL level $f_{15}$.

The canonical average for a quantity $X$ can be calculated as
\begin{equation}\label{mean}
 \bar{X}(\beta,\epsilon_H,\mu)=\dfrac{\sum_{E_H, N}\langle X\rangle_{E_H, N} g(E_H, N) 
  e^{-\beta (\epsilon_H E_H - \mu N)}}{\sum_{E_H, N} g(E_H, N) e^{-\beta (\epsilon_H E_H -\mu N)}} ,
\end{equation}
where $\langle X\rangle_{E_H,N}$ is the microcanonical average accumulated during the simulation.

With the density of states in hands, we can calculate any thermodynamic quantity of the system for different values of the energy parameters and temperature, among which we highlight the canonical probability, Eq. \ref{probability}, which can be rewritten in terms of the joint density of states like
\begin{equation}\label{eq:prob}
 P(E_H, N,T)=\dfrac{ g(E_H, N)e^{-\beta (\epsilon_H E_H - \mu N)}}{\sum_{E_H, N} g(E_H, N) e^{-\beta (\epsilon_H E_H -\mu N)}}. 
\end{equation}
This quantity indicates the probability of a configuration with hydrogen bonding energy $\epsilon_H E_H$ and chemical potential energy $\mu N$ happening, at a temperature $ T $, with $\epsilon_H $ and $\mu$ fixed. In a continuous phase transition, $P(E_H,N,T)$ has a gaussian shape and, as $T$ increases, its goes from an acute width for $T<T_c$ to a wider one at $T\sim T_c$, returning to its original shape for $T>T_c$.

The Wang-Landau algorithm performs a random walk through the energy space independent of temperature, making it impossible to obtain the configurations as a function of temperature. In the Metropolis algorithm\cite{Metropolis1953}, obtaining the configurations for a given temperature is a simple task, since the acceptance probability depends on the temperature. However, these configurations may be conditioned to the path taken by the algorithm, showing only some of the possible configurations probable for that temperature. This fact can influence the description of the phase transition.

Therefore, we propose an algorithm based on WL to obtain the configurations as a function of temperature, using the canonical probability to identify the values of $E_H$ and $N$ that maximize it, being possible to filter the configurations on the random walk. The only requirement of the algorithm is to know the density of states $a~priori$, being calculated from a previous simulation of WL.
\begin{figure*}
\centering
\includegraphics[scale=0.3,angle=-90]{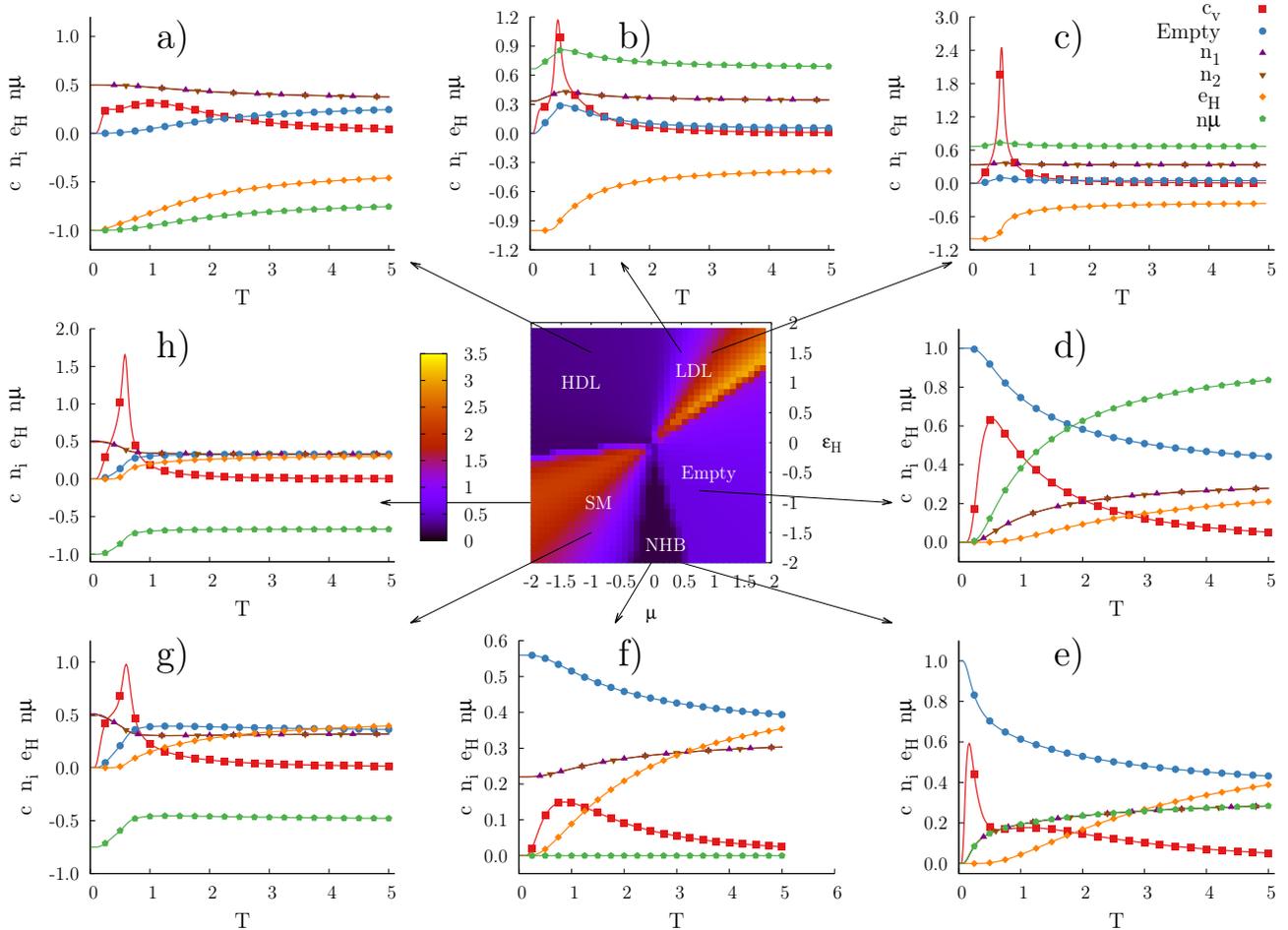}
\caption{Phase diagram and thermodynamic properties for the Bell-Lavis model using $\epsilon_{vdW}=0$ for $L=12$. The maximum of the specific heat is represented in the center depending on
parameters $\epsilon_H$ and $\mu$. Around it are represented the specific heat, the densities of empty sites, type I molecules, and type II molecules, the energy density of the chemical potential and the density of hydrogen bonds for each of the regions delimited by the phase outlines, the different colors shown in the figure. The values of $(\mu:\epsilon_H)$ used to obtain the thermodynamic properties were a) $(-1.0:1.0)$, b) $(0.5:1.0)$, c) $(1.0:1.0)$, d) $(1.5:-1.0)$, e) $(0.5:-2.0)$, f) $(0.0:-1.5)$ g) $(-0.75:-1.5)$ and h) $(-1.0:-1.0)$.
\label{fig:painel}}
\end{figure*}

To obtain the configurations that maximize the canonical probability, we apply the following algorithm:\begin{enumerate}
\item Set values for $T_0$, $\epsilon_H$ and $\mu$.

\item Adopt a previously simulated joint density of states. 

\item Calculate the number of hydrogen bonds and the number of particles that maximize the energy probability, $E_H^{max}$ and 
$N^{max}$, using Eq. \ref{eq:prob}

\item Choose a new configuration by randomly changing the state of a lattice site, whose acceptance probability is given by
\begin{equation}
p(E_H, N \rightarrow E_H', N')=\text{min}\left\{\frac{\tilde{g}(E_H, N)}{\tilde{g}(E_H', N')},1\right\},
\end{equation}
where $\tilde{g}$ is the previously simulated density of states. 
\label{loop}

\item If $E_H$ and $N$ are equal to $E_H^{max} $ and $N^{max}$ of the maximum probability, save the configuration and finish the algorithm. If not, return to step \ref{loop}.  
\end{enumerate}

This algorithm allows us to obtain all the most probable configurations, even if they are separated by a potential barrier, as is the case of a discontinuous transition, where the energy probability presents a valley between the two most probable configurations.

\section{Results}

With the joint density of states, we can obtain the thermodynamic quantities for any values of $\epsilon_H$, $\mu$, and the temperature, favoring the access to a huge amount of information. To facilitate the visualization of the system behavior, we have drawn up a diagram in which we present the maximum of the specific heat as a function of $\epsilon_H$ and $\mu$, which is shown in the center of Fig. \ref{fig:painel}. We are considering that changes in the system phases are accompanied by an increase in energy fluctuations. In the diagram, one can see that there are different regions, such that in each of them the thermodynamic properties have the same characteristics. Around this diagram, we show some of the thermodynamic properties for each region, and then, we name the regions concerning the phase present in the ground state. The abbreviation $NHB$ means no hydrogen bonds, while $SM$ stands for single-molecule.

In the $HDL$ phase shown in Fig. \ref{fig:painel} a), the density of empty sites is zero, and the particle densities show that the lattice possesses 50\% of molecules of each type, at low temperatures. The specific heat has two smooth maxima, with the first related to the breakdown of hydrogen bonds and the second to the release of particles from the lattice. This can be seen in the behavior of the energy density of the chemical potential and the energy density of the hydrogen bonds, which initiate a slight increase at different temperatures.
\begin{figure*}
\includegraphics[scale=0.16]{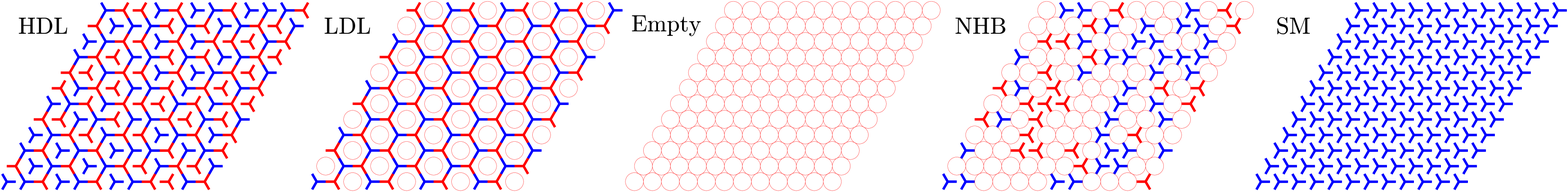}
\caption{Phases found in the different regions of the phase diagram: $HDL$,  $LDL$,  Empty, $NHB$ e $SM$  \label{fig:config}}
 \end{figure*}
 
In the $LDL$ phase shown in Fig. \ref{fig:painel} b) and c), the particle densities and the empty site density start with the same value $1/3$. The specific heat has a prominent maximum in the lightest region and two peaks in the darkest section. The first, the softest, is related to the entry of particles into the lattice without breaking the hydrogen bonds. The second maximum is associated with the breakdown of the hydrogen bonds and is more pronounced than the first. The energy density of the chemical potential shows a maximum around the temperature of the specific heat peak, showing that in this region, we have a maximum in the particle density. This behavior suggests the presence of a transition from a $LDL$ phase to the $HDL$. The intensity of the maximum specific heat in this region is related to the abrupt change in the hydrogen bond energy density.

Fig. \ref{fig:painel} d) shows the behavior of the quantities for the region that we are calling empty, in which at low temperature, we have the lattice empty. This happens due to the signal of the chemical potential that repels the particles from the lattice sites. With increasing temperature, the particle entry favors the variation of energy related to the chemical potential causing the maximum of the specific heat. This increase, in principle, does not form hydrogen bonds, which only starts to be relevant for temperatures above the temperature of the maximum of the specific heat. The formation of hydrogen bonds occurs so smoothly that the second peak in heat specific is imperceptible.

At the interface between the empty and non-hydrogen bonded phases 
($NHB$), Fig. \ref{fig:painel} e), we see the presence of two maxima. The first is more acute due to the abrupt entry of particles, and the second is smooth due to the hydrogen bonds that start to happen. It is important to note that the first peak occurs at extremely low temperatures, disappearing in the $NHB$ region, Fig. \ref{fig:painel} f), whose ground state is formed by molecules of types I and II interspersed with empty sites that inhibit, at low temperatures, the formation of hydrogen bonds.

Finally, in the $SM$ region shown in Fig. \ref{fig:painel} g) and h) the configurations at low temperatures are two, one with all sites with type I molecules or fulfilled with type II molecules. In this region, we have two distinct colors that are associated with the maximum of the specific heat, which in the less intense section, Fig. \ref{fig:painel} g), presents two peaks, and in the most intense region, Fig. \ref{fig:painel} f), a single maximum resulting from the merging of the previous two. These maxima are related to particle output and the formation of hydrogen bonds.

The identification of the ground state configurations for each region was complemented by the acquisition of the configuration that maximizes the canonical probability at low temperatures. The result of this procedure can be seen in Fig. \ref{fig:config}. Analyzing these configurations, one can see that the $LDL$ and empty phases have no degeneration. The $SM$ phase is doubly degenerate and, the $HDL$ and $NHB$ phases are highly degenerate.
\begin{figure}[b]
\centering
\includegraphics[scale=0.6,angle=-90]{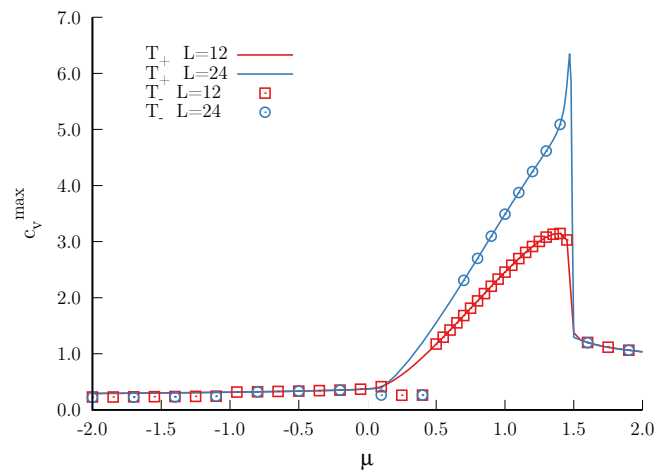}
\caption{Cut of the phase diagram for the maximum of the specific heat at $\epsilon_H = 1$ and $\epsilon_ {vdW}=0$ for $L=12$ and 
$L=24$. The line represents the maximum for the highest temperature and the circles and crosses for the lowest temperature. \label{fig:fss}}
\end{figure}

To carry out a more detailed analysis, we will restrict our study to $\epsilon_H=1$. We are choosing this range due to the possibility of comparing our results with previous works and because it contains the region that has the maximum particle density.

\subsection{System behavior for $0<\mu<1.5$}

To identify indications of finite-size effects in the specific heat, for $\epsilon_H=1$ and $-2<\mu<2$, we compared the maxima of specific heats for sizes $L=12$ and $L=24$, which results can be seen in Fig. \ref{fig:fss}. We realized that the only region that has a  finite-size effect in the specific heat is $0<\mu<1.5$. It does not mean that in the other regions it cannot have phase transitions, they are just not observed by the specific heat. The critical exponents for this region were obtained by Monte Carlo simulations using Metropolis algorithm\cite{Simenas2014}, where they used two parameters of antiferromagnetic order, staggered magnetization and the difference between the particle and empty site densities. They were interested in the transition between a $LDL$ configuration with some molecules inside the hexagons and the disordered one, which is characterized by the sharpest maximum of the specific heat. But, before we go-ahead to a finite-size scaling analysis, let us look at the behavior of the system in the region where this phenomenon occurs for the specific heat.
\begin{figure}[t]
\centering
\includegraphics[scale=0.38,angle=-90]{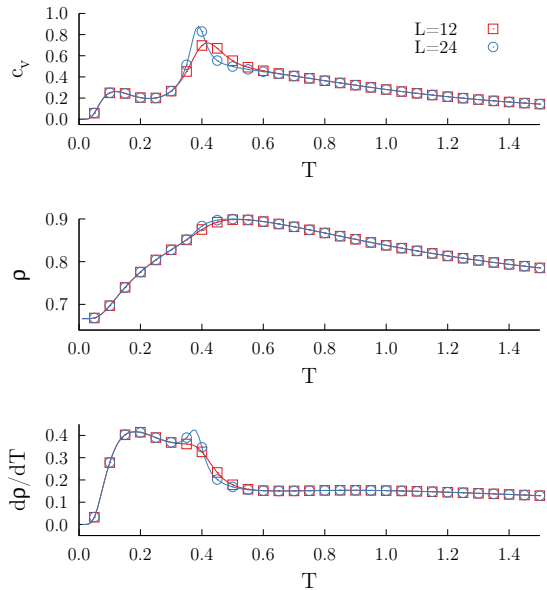}
\caption{Specific heat, particle density and particle density fluctuation for $L=12$ and $L=24$ with $\epsilon_H=1$ and 
$\mu=0.3$. \label{fig:cv-mu}}
\end{figure}

In this region, Fig. \ref{fig:cv-mu}, we show the behavior of  specific heat, particle density and particle density fluctuation for $L=12$ and $L=24$ with $\epsilon_H=1$ and $\mu=0.3$. We see in the specific heat the presence of two maxima, and the finite-size effect occurs only in the second. When we look at the particle density, we notice that there are a smooth maximum and a finite-size effect next to it. When we calculate the particle density fluctuation, we see the presence of three maxima, the last, between $T=0.6$ and $1$ is quite smooth. Such behavior indicates that in the particle density curve, we have three plateaus, suggesting the occurrence of three transitions. It seems that only the second transition has a finite-size effect. This situation is quite different from those found in the standard works that deal with phase transitions in spin systems, where it is well known which transition is taking place. 
\begin{figure}[t]
\centering
\includegraphics[scale=0.19,angle=0]{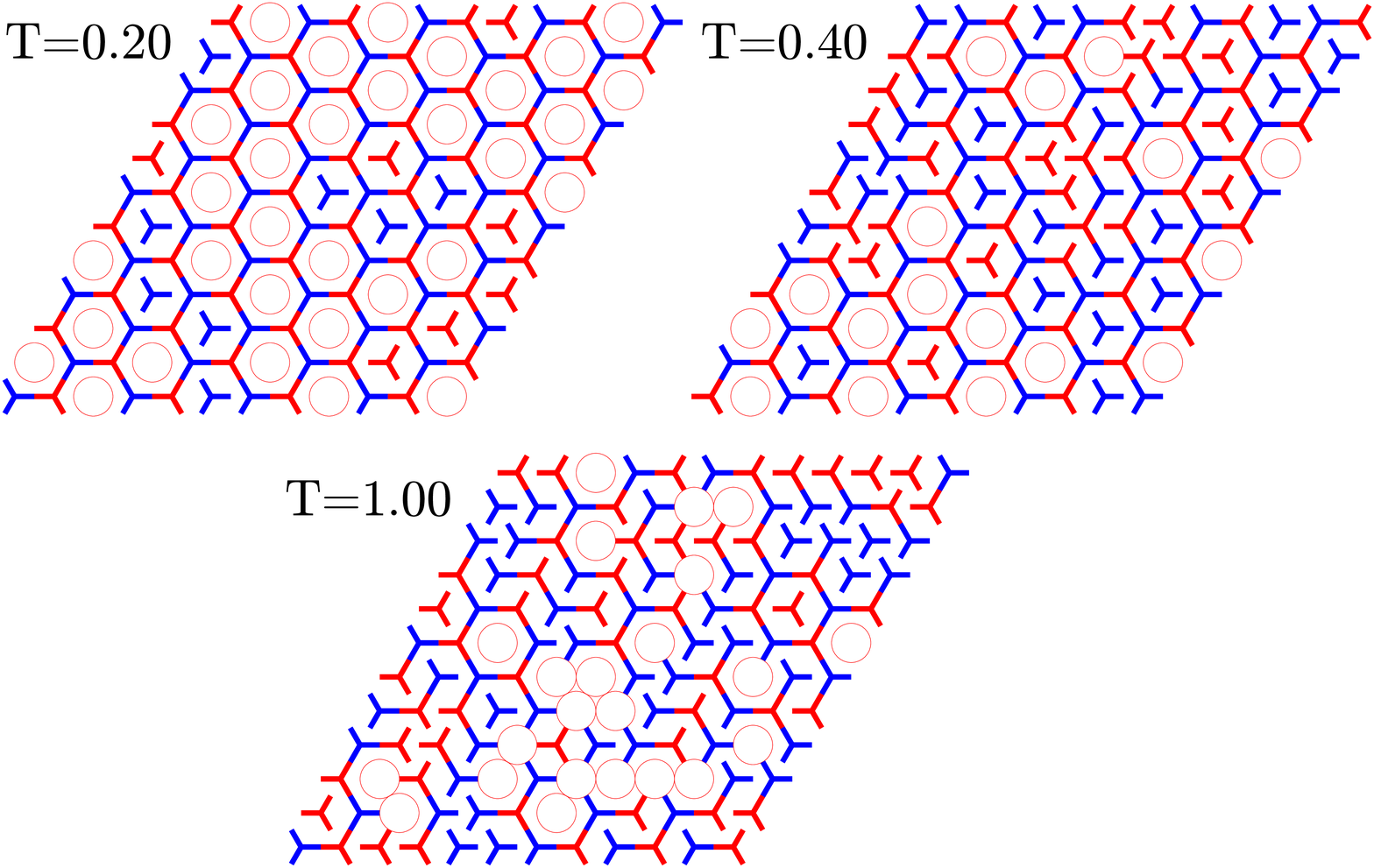}
\caption{Configurations that maximize the energy probability for $\epsilon_H=1$ and $\mu=0.3$ with $L=12$. \label{fig:tresconfig}}
\end{figure}

To understand what may be happening in the region of the three maxima, we will analyze the configurations that maximize the canonical probability for $\epsilon_H=1$, $\mu=0.3$, and $L=12$. In Fig. \ref{fig:tresconfig} we show the configurations for  $T=0.2,~ 0.4$ and $1.00$ corresponding to the maxima of the particle density fluctuation. Here the ground state is the $LDL$ configuration, and as the temperature increases, at $T=0.2$, the system undergoes a transition to a $HDL_0$, as shown in Fig. \ref{fig2} b), both phases start to coexist. Increasing the temperature a little more, $T=0.4$, we see the appearance of other $HDL$ phases that start to coexist with the previous two. Finally, this mixture of phases gives way to the disordered phase.

Observing these configurations, we can assume that the first transition occurs from a $LDL$ to $HDL_0$ phase. However, both phases coexists, highlighted by the first maximum in specific heat. Its behavior is known as a Schottky anomaly, having no scaling effect. The finite-size effect observed on the fluctuation of the particle density, signaling the appearance of different $HDL$ phases such as the zig-zag (as shown in Fig. \ref{fig2} c)), which also starts to coexist with the previous ones. This phase has no long-range order and, the number of phases coexisting can be confused with a disordered phase. However, when we look at the phase that characterizes the third maximum in the particle density fluctuation, we see a breakdown of the order mentioned previously, showing that now the phase is in fact disordered. We also emphasize that in these configurations there are bubbles, e.g., clusters of empty sites.

\begin{figure}[t]
\centering
\includegraphics[scale=0.38,angle=-90]{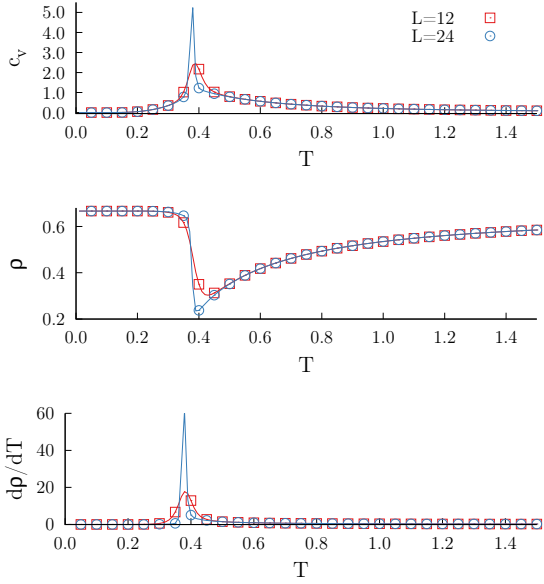}
\caption{Specific heat, particle density and particle density fluctuation for $L=12$ and $L=24$ with $\epsilon_H=1$ and 
$\mu=1.48$. \label{fig8}}
\end{figure}

By increasing the chemical potential, the three transitions occur at temperatures closer to each other, increasing the energy fluctuation and the maximum of the specific heat. At $\mu\sim1.5$, the specific heat, the particle density, and the particle density fluctuation show typical phase transition behavior found in standard models, as can be observed in Fig. 
\ref{fig8}, where we show the results for $\mu=1.48$. We noticed that the behavior of the particle density changed dramatically, not showing the maximum, indicating the absence of transition $LDL$-$HDL$. Now what we see is an abrupt decreasing in particle density around $T=0.4$, and then a smooth increase for $\rho=2/3$. This behaviour starts at $\mu\sim1.3$. Since the maximum of the specific heat does not show any signal for such behaviour, we have to scrutiny the particle density changes while testing several values of $\mu$. 

\begin{figure}[t!]
\centering
\includegraphics[scale=0.19,angle=0]{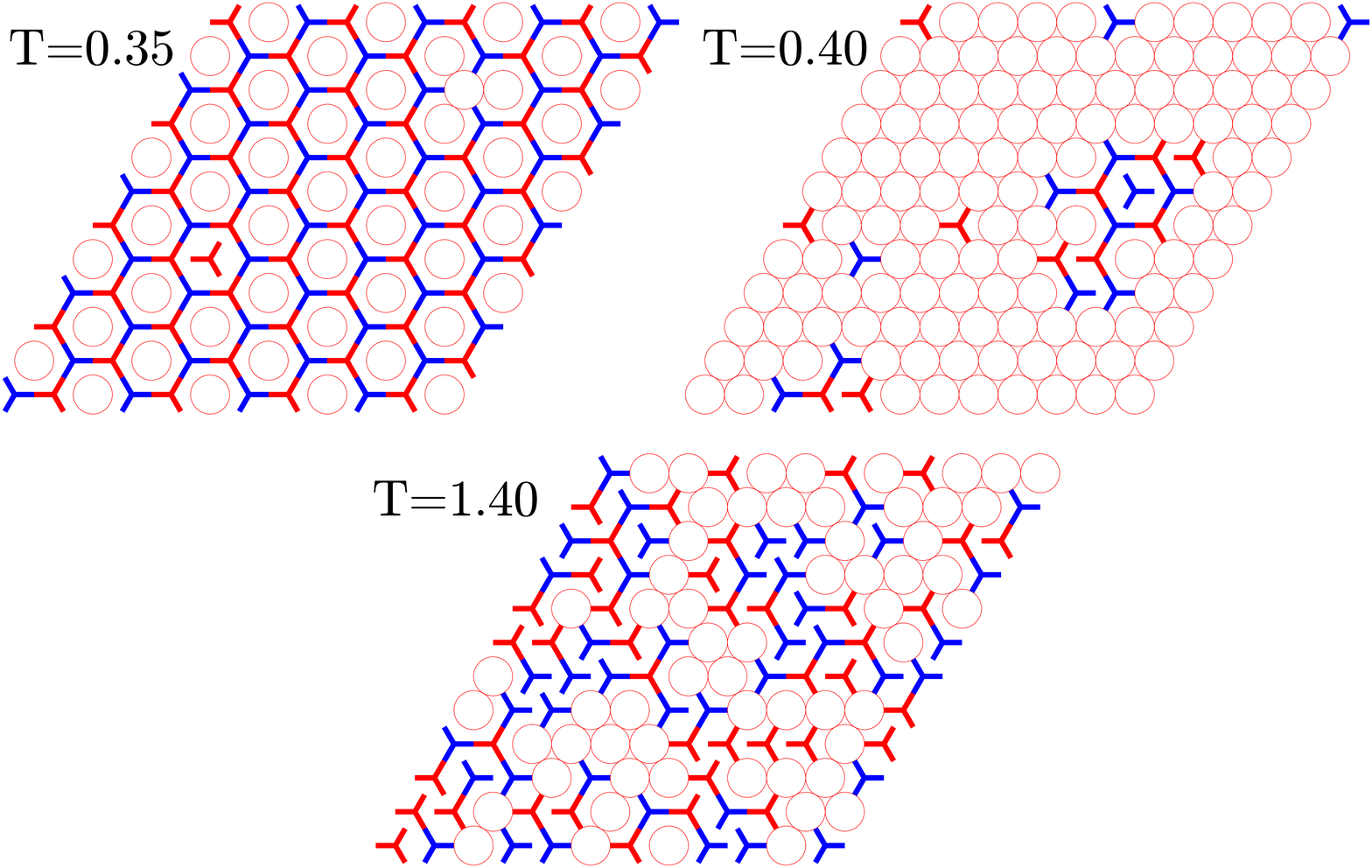}
\caption{Configurations that maximize $P(E_H,N,T)$ for $\epsilon_H=1$ and $\mu=1.48$ with $L=12$. \label{fig9}}
\end{figure}

The change in particle density is evident when we look at the configurations around $T=0.4$, Fig. \ref{fig9}. At $T=0.35$, the system still has the characteristics of the $LDL$ ground state with a few particles inside the hexagons and a few more empty sites than the conventional one. At $ T=0.4$, the configuration is different, being formed mostly by empty sites. Molecules that remain in the lattice tend to form hydrogen bonds. This abrupt variation is indicative that a discontinuous transition is taking place.

At $\mu=1.5$, the system presents a degenerated ground state, composed by the $LDL$ configuration and the empty lattice. Here the two phases do not coexist and the system is in one or the other. However, both have the same probability of being found in the ground state, contributing equally to the mean density of particles, making it start at $\rho=1/3$. For $\mu>1.5$, the ground state is the empty lattice.

\subsection{Finite-size analysis}

So far, we have only discussed the behavior of the system by analyzing which configurations are present in each region and at the temperatures of the specific heat maxima and the fluctuation of the particle density. No analysis of critical exponents was performed.
\begin{figure*}
\centering
\includegraphics[scale=0.6,angle=-90]{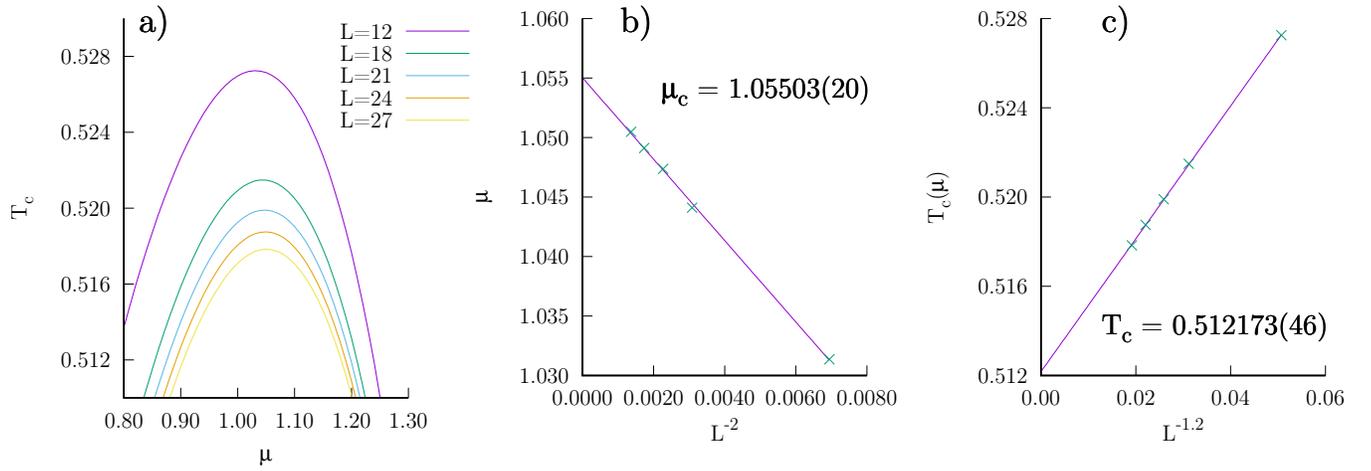}
\caption{a) Peaks in the $T_c$ x $\mu$ phase diagram. b) linear fitting for the critical potential at the maximum of the phase diagram and c) linear fitting for the critical temperature at the maximum of the phase diagram. \label{fig10}}
\end{figure*} 

As we increase the chemical potential, from $\mu=0$, the temperature of the peak of the specific heat increases until reaching a maximum, Fig. \ref{fig10} a), which indicates the coexistence among the 
$LDL$, $HDL$ and empty phases. After the maximum in the temperature diagram, we do not see the presence of the $(LDL ~ HDL_0)-HDL$ transition. To identify this maximum, we assume that the chemical potential obeys the power law given by 
\begin{equation}\label{potencialcritico}
\mu (L)=\mu_c+bL^{-2},
\end{equation}
where $\mu(L)$ is obtained from the maximum of the plot of $T_c$ as a function of $\mu$ for each lattice size, $\mu_c$ the chemical potential at the thermodynamic limit and $b$ is a constant. The result of the linear fitting for $L=12,~18,~21,~24$ and $27$ is shown in Fig. \ref{fig10} b), whose chemical potential value for the coexistence of the three phases is $\mu _c=1.05503(20)$. This value is close to the one reported by the reference \cite{Simenas2014}, for which the maximum susceptibility to staggered magnetization has a minimum as a function of chemical potential. To determine the temperature of the coexistence of the $LDL,~HDL$ and empty phases, we assume that the maximum temperature of the $T\times\mu$ diagram is asymptotic at
\begin{equation}
T_c(L)=T_c+dL^{-1/\nu},
\end{equation}
where $d$ is a constant and $T_c(L)$ is the peak of curve $T_c$ as function $\mu$. The result of this fitting is shown in the Fig. \ref{fig10} c), getting $T_c=0.512173(46)$.

The order parameter must reflect the ordered state through a physical quantity that ranges between zero in the disordered phase and nonzero in the other. In the region $0<\mu<1.05503$, the system has several phase coexistences, making it difficult to define such that parameter. Simenas {\textit et al.} \cite {Simenas2014} concluded that the phase transition occurred from the $LDL-HDL_0$ coexistence to the disorder, however, as shown, the maximum in specific heat only presents a finite size effect when the transition from coexistence between $LDL$ and $HDL_0$ occurs to the $LDL-HDL$ coexistence. It is clear that at that time, they only had the information of the order parameter named staggered magnetization, making it impossible to identify the transition that was taking place, a fact that does not refute the results found but must be understood from a new point of view.

\begin{figure*}
\centering
\includegraphics[scale=0.6,angle=-90]{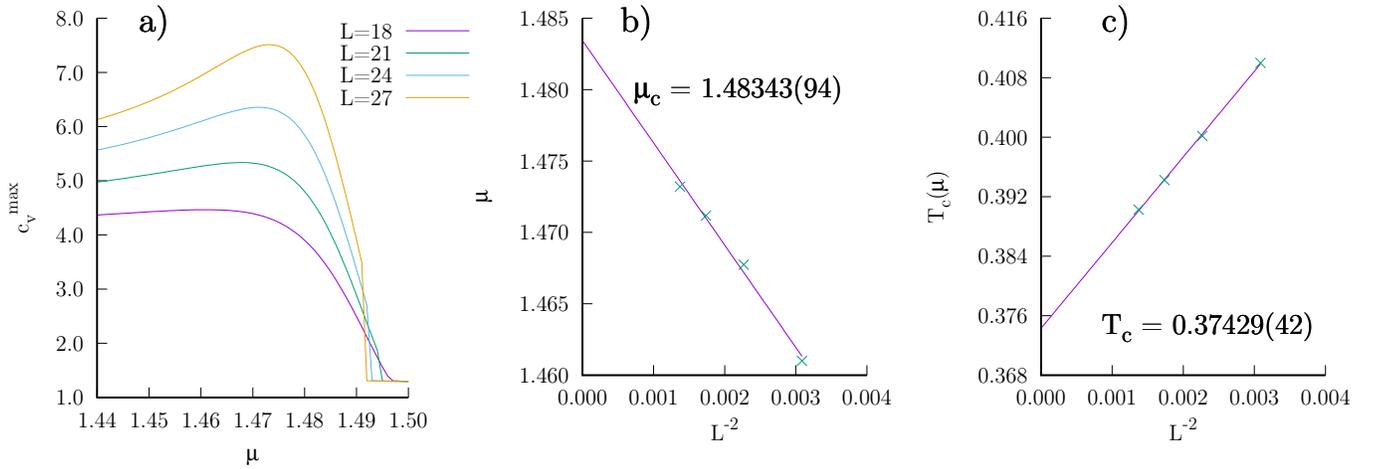}
\caption{a) Peaks in the $c_v^{max}$ x $\mu$ phase diagram. b) linear fitting for the critical potential at the maximum of the phase diagram and c) linear fitting for the critical temperature at the maximum of the phase diagram.  \label{fig11}}
\end{figure*} 
From $\mu>1.05503$, we start to observe significant changes in the behavior of particle density, showing that a phase with empty sites is emerging while increasing the temperature and the closer to $\mu=1.5$ while approaching the new phase. In this region, we expect three phases to coexist: $LDL$, $HDL_0$, and the empty one. It is not yet clear how we will distinguish the transition to the disordered state. However, what we can say is that the coexistence between the three phases disappears for $\mu\sim1.48$, leaving only the transition from an $LDL$ phase to an empty one, as can be seen in Fig. \ref{fig9}. Besides, as the temperature rises, the system moves into a disordered configuration, as can be seen in Fig. \ref {fig9} $T=1.40$.

To identify the temperature and chemical potential at which the  $LDL-HDL_0$ transition ceases to occur, we observe the behavior of the peak of specific heat as a function of the chemical potential for $L=18,~ 21,~ 24$ and $27$, represented in Fig.  \ref{fig11} a). We realize that it has a maximum that we suppose to scale according to the power-law given by Eq. \ref{potencialcritico}. The result of the linear fitting can be seen in Fig. \ref{fig11} b), whose critical chemical potential is $\mu _c=1.48024(50)$. To determine the critical temperature, we use the peak temperature of the maximum specific heat that we assume to scale according to
\begin{equation}
T_c(L)=T_c+dL^{-2},
\end{equation}
where $d$ is a constant. The result of the fitting for the critical temperature is shown in Fig. \ref{fig11} c), whose value is $T_c=0.3791(10)$. The values of $T_c$ and $\mu _c$ are in agreement with those found in Ref. \cite{Simenas2014}. Now the two scaling laws are for a discontinuous transition.

\begin{figure}[t]
\centering
\includegraphics[scale=0.6,angle=0]{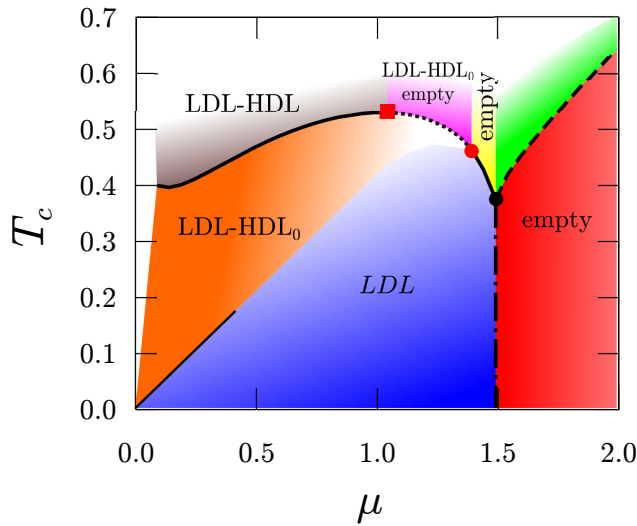}
\caption{Schematic phase diagram $T_c\times \mu$ in the region where the finite-size effect occurs. The thicker continuous line represents a continuous transition, the thinner one only demarcates the position of the maxima of the first peak of the specific heat, which is related to the $LDL-HDL_0$ transition.   \label{fig_diagrama}}
\end{figure} 

To summarize all the information obtained by analyzing the configurations and the finite-size scaling study, we made a $T_c\times\mu$ schematic diagram that is shown in Fig. \ref{fig_diagrama}. One can see that it has several phase coexistences. Unlike what is observed in the magnetic model, these coexistences are not limited to a critical line, but they occupy an area of the phase diagram. We use lines to delimit transitions between different coexistences. The first to be cited here is the line that divides the $LDL$ phase from the $LDL-HDL_0$ coexistence. It derives from the temperature of the first maximum of the specific heat and its continuation is made impossible by the fusion of its two maxima. However, we can extrapolate the trend until we find another line of transition, seen by the color difference. This line has no finite-size effect, which is why we have not assigned a phase transition order to it.

The transition between $(LDL-HDL_0)-(LDL-HDL)$ is second order and the critical exponents were obtained by Ref. \cite{Simenas2014}. It ends at the point, $\mu _c=1.05503(20)$ and $T_c=0.512173(46)$, marked by a square where the phases $LDL~, HDL$, and empty coexist. The dotted line marks the transition $(LDL-HDL_0)-(LDL-HDL_0-$empty). The order of this transition is one of the targets of future studies since we have the characteristics of two transitions occurring. This type of transition has already been reported in the Baxter-Wu model spin-1 \cite{Jorge2019}. The line ends at $\mu _c=1.48024(50)$ and $T_c=0.3791(10)$, marked by a circle, where we begin the $LDL-$empty phase transition, which is first order and ends at $\mu=1.50$.

The line interrupted by dots marks a first-order transition that occurs when maintaining the temperature constant and varying the chemical potential. It has different characteristics from the 
$LDL-$empty transition found with increasing temperature. Finally, the broken line delimits the empty phase of a phase partially filled by molecules. We are unable to observe a pattern of any phases in this region so that the transition takes place from an empty phase to the disorder.

\section{Conclusions}
In this work, we investigated the Bell-Lavis model using entropic simulations for different values of the energy parameters. We have shown, through the phase diagram, the behavior of the thermodynamic quantities in the five regions, comprising $HDL$, $LDL$, empty, 
$NHB$ and $SM$, corresponding to the ground state in each of them. Further analysis showed that for $\epsilon_H=1$, the only region that has a finite size effect is $0<\mu<1.5$. It was divided according to the behavior of the particle density. For $0<\mu<1.05503$ the particle density has a maximum, caused by the transition between the $LDL-HDL_0$ and $LDL-HDL$ coexistences. For $1.05503<\mu<1.48024$ the particle output from the lattice causes the maximum particle density to disappear, giving rise to a minimum, deeper the greater the chemical potential. In this region, the observed transition is between $LDL-HDL_0$ and $LDL-HDL_0-$empty coexistence, with a discontinuous and continuous transition characteristic. In the narrow region $1.48024<\mu<1.5$ the discontinuous transition occurs from a $LDL$ phase to the empty lattice, without coexistence between the phases. We estimate the value of the chemical potential and temperature for which we have the coexistence $LDL-HDL_0-HDL-$empty yielding $\mu _c=1.05503(20)$ and $T_c=0.512173(46)$. This point marks the end of the $LDL-HDL$ coexistence and the beginning of a $LDL-HDL_0-$empty coexistence. We also estimated the point where 
the $LDL-HDL_0-$empty coexistence ends up obtaining 
$\mu _c=1.48024(50)$ and $T_c=0.3791(10)$. At $\mu=1.5$ the $LDL$ and empty phase have the same probability at low temperatures, however, they do not coexist. The transition occurs from the $LDL$ or empty phase to the disorder. This situation is the same as in the 2D Ising model, where the positive ferromagnetic phase has the same probability as the negative ferromagnetic phase.

We note that the transition to disorder occurs at a temperature higher than the transition temperatures observed in this work and that the quantities analyzed here do not show this transition. However, for $\mu>1.5$, the transition occurs from the empty phase to the disorder and has no finite size effect.

Regarding the order parameter to observe the transition between the $LDL$ and $HDL$ phases, we see that the particle density and the specific heat can describe this transition, suggesting that the results are given by Ref. \cite{Simenas2014} should be reinterpreted.

Finally, we emphasize the need to visualize the most probable configurations for a set of energy and temperature parameters to determine the phases involved in the phase transition. And that the algorithm proposed in this work allowed access to such configurations, corroborating results obtained previously.

\begin{acknowledgments}

We acknowledge the computer resources
provided by LCC-UFG. L. S. F. acknowledges the support by CAPES.
\end{acknowledgments}

\section*{Data availability}
The data that support the findings of this study are available from the corresponding author
upon reasonable request.

\bibliography{referencias.bib}

\end{document}